# Controlled transfer of transverse optical angular momentum to optically trapped birefringent particles

Alexander B. Stilgoe,[*] Timo A. Nieminen, and Halina Rubinsztein-Dunlop

*School of Mathematics and Physics,*

*The University of Queensland,*

*St. Lucia, Brisbane, QLD, 4072,*

*Australia and*

*Australian Research Council Centre of*

*Excellence for Engineered Quantum Systems,*

*The University of Queensland,*

*St. Lucia, Brisbane, QLD, 4072,*

*Australia*

**ABSTRACT**

We report on the observation and measurement of the transfer of transverse angular momentum to birefringent particles several wavelengths in size. A trapped birefringent particle is much larger than the nano-particles systems for which transverse angular momentum was previously investigated. The larger birefringent particle interacts more strongly with both the trapping beam and fluid surrounding it. This technique could be used to transfer transverse angular momentum for studies of diverse micro-systems. Thus, it can be used for investigation of the dynamics of complex fluids in 3D as well as for shear on cell mono-layers. The trapping of such a particle with highly focused light is complex and can lead to the emergence of effects such as spin–orbit coupling. We estimate the transfer of spin angular momentum using Stokes measurements. We outline the physics behind the construction of the beam used to control the particles, perform quantitative measurement of transverse spin angular momentum transfer, as well as demonstrate the generation of fluid flow around multiple rotation axes.

**INTRODUCTION**

The conservation of momentum is a fundamental principle in non-relativistic physics. As a consequence of the energy a photon carries, there is a corresponding momentum to its movement in space. It can transfer momentum to matter through absorption and scattering [1–5]. An object scattering light experiences an impulse from the field through the conservation of momentum. Optical tweezers trap particles in 3D through linear momentum transfer from strongly focused light [2, 6–8]. Optical forces have found great use in analyzing biological systems in terms of mechanical components [9] and contributed to experiments on exotic states of matter such as BEC [10]. Later experiments demonstrated optical angular momentum transfer to trapped particles [5] which was a great development in controlling and characterizing microscopic systems [11–14]. Angular momentum transfer is usually restricted to the axis defined by beam propagation [15–17]. This is not an absolute, and more recently the transfer of angular momentum around an axis orthogonal to the beam propagation direction was demonstrated on absorbing particles using retro-reflected laser beams of non-uniform circular polarizations [18, 19]. These investigations have led us to believe that trapping and control in 3D over larger bireferingent particles is possible. Significantly more angular momentum can be transferred to larger particles and thus they can perform work on the surrounding environment as microscopic motors. The concept behind this is shown in Fig. 1a). A bespoke beam depicted as propagating from the bottom of the page interacts with the larger birefringent particle to induce spinning and rotation about the axis denoted with the curved arrow.

We report on the observation and measurement of the transfer of transverse angular momentum to birefringent vaterite particles (positive uni-axial crystals) several wavelengths in size and trapped with optical tweezers in a transmission optical system. The trapping of such a particle requires highly focused structured light which can lead to the emergence of effects such as spin–orbit coupling [20–23] which effects the measurement of particle dynamics. We estimate the transfer of spin angular momentum using Stokes measurements. However, we also show that these measurements cannot yield the actual angular momentum transferred in this system due to the coupling. We discuss the difficulties and prospects for quantitative measurement of angular momentum that this system poses.

In addition to the studies of fundamental light-matter interactions discussed here, this technique could be used to transfer transverse angular momentum for studies of diverse micro-systems. A trapped birefringent particle, such as vaterite [12], is much larger than the nano-particles systems where transverse angular momentum was previously investigated [18, 19, 24]. The larger birefringent particle interacts more strongly with both the trapping beam and fluid surrounding it. Thus, it can be used for investigation of the dynamics of complex fluids in 3D as well as for shear on cell mono-layers.

**OPTICAL MOMENTUM TRANSFER**

It is the transport of momentum by light that allows light to be used to trap and manipulate particles, such as with optical tweezers. The momentum flux **p** is related to the energy flux $P$ by the speed of propagation in the medium:

$$\mathbf{p} = P|\mathbf{k}|/(nc) \tag{1}$$

where **k** is the wave vector, $n$ is the refractive index of the medium, and $c$ is the speed of light in free space. The forces responsible for radial trapping result from deflection of the trapping beam. If the particle in the trap deflects the beam to the right, then conservation of momentum demands that a reaction force acts on the particle to the left.

The origin of the axial force opposite to the direction of propagation that is necessary for three-dimensional trapping is less obvious. The key point is that a collimated beam (or a plane wave, or a ray) of a given power has a higher momentum flux than a converging or diverging beam. This is because the local wavevector is always parallel to the direction of propagation in a collimated beam while it is at an angle to the direction of propagation for a converging or diverging beam (Fig. 1b) in which case only the vector component in the propagation direction contributes to the total momentum flux of the beam. If the trapped particle changes the convergence or divergence of the beam, it changes the component of the momentum flux in the axial direction. If it makes the beam more collimated (i.e., less convergent or divergent), it increases the momentum of the beam in the direction of propagation, resulting in a restoring force acting on the particle in the opposite direction, as required for three-dimensional trapping against axial forces resulting from reflection and absorption [2].

Light can also carry angular momentum and can be used to control the orientation of particles and to rotate them. An important point about the angular momentum of light is that it can be either *spin* or *orbital* angular momentum. We can write the total angular momentum density as the sum of spin and orbital angular momentum densities:

$$\mathbf{j} = \mathbf{l} + \mathbf{s}, \qquad (2)$$

where $\mathbf{l}$ is the orbital angular momentum density and $\mathbf{s}$ is the spin angular momentum density. The key theoretical difference between spin and orbital angular momenta is that the spin angular momentum density is independent of the choice of coordinate system [25, 26]. Thus, a beam of light can carry spin angular momentum about the beam axis on the beam axis itself, since no moment arm is required. The spin and orbital angular momentum fluxes of a beam of light can be found from the densities by integrating across the cross-section of the beam. If the beam is paraxial and monochromatic (or at least quasi-monochromatic), the circular polarization of the beam simply determines the spin angular momentum: $\hbar$ per photon for left-circular polarization and $-\hbar$ per photon for right-circular.

If the beam is non-paraxial the field structure becomes specific to particular regions of space. While a non-paraxial beam cannot be unambiguously described by a single polarization like a paraxial beam, the far field of the beam—which is locally of the form of a plane wave—can be locally described by a polarization. (This also means that the spin angular momentum can be measured by determining the Stokes parameters of the light in the far field, while typical schemes for the measurement of orbital angular momentum require information about the phase of the light [27].)

If the light is everywhere locally left-circularly polarized, then the spin angular momentum density is $\hbar$ per photon in the direction of the local wavevector. Only the component in the direction of the beam axis will contribute to the total spin angular momentum of the beam. While a collimated beam will have a total spin of $\hbar$ per photon, a converging beam must have less spin, as shown in Fig. 1c. This is the same geometric reasoning that shows that a converging or diverging beam has a lower linear momentum flux than a collimated beam (Fig. 1b).

However, this reduction only applies to the spin angular momentum flux. If the converging beam is produced (as usual) by focussing a collimated beam using a rotationally symmetric lens, the angular momentum per photon will be unchanged [28, 29]. Since the spin angular momentum is reduced, an equal amount of orbital angular momentum is introduced [20, 25, 30], as shown in Fig. 1c. This spin–orbit conversion of angular momentum can have measurable consequences, as we will see below.

A key difference between the behaviour of the spin density and the linear momentum density is that the spin density can be anti-parallel to the local wavevector, i.e., $-\hbar$ per photon. Thus the axial component of the spin flux of a complex beam need not always be in the same direction, and it is possible to engineer a beam such that the axial components of the spin flux cancel across the beam, and the beam has a total spin angular momentum about the beam axis of zero. It is also possible to produce a quite abnormal condition for a beam of light: a non-zero spin flux about an axis *normal* to the beam axis (Fig. 1d). Such transverse spin is a potential pathway to the rotation of particles in optical traps about axes normal to the beam axis.

For a practical realisation of this, we must consider something more concrete than the unspecified "beam" above and a Gaussian beam is the obvious starting point, though vectorial polarization shaping of the light is also possible [31, 32]. The superposition of two non-co-linear Gaussian beams of opposite circular polarization of equal power and both at the same angle to the common beam axis will result in zero total axial spin and maximum transverse spin [18]. If both beams are focused using a high-numerical-aperture lens, they can form an optical trap. Noting that in a microscope system, changing the angle of a beam is equivalent to changing its position at the back aperture of the main focusing optics, the two oppositely-polarized beams can be initially parallel and directed onto opposite sides of the back aperture of the microscope objective.

A simple way to transfer the transverse spin angular momentum to a trapped particle is to use absorption. However, absorption forces inherently limit the maximum size of the particle that can be trapped in optical tweezers, since absorption will also transfer linear momentum, acting to push the particle out of the trap [33]. Thus, direct transfer of transverse spin angular momentum has previously been restricted to nanoparticles [19].

Birefringent vaterite microspheres are commonly used for the transfer of axial spin angular momentum in optical traps [12, 14, 34] In this case, they act approximately as waveplates, and change the degree of circularly polarization of the trapping beam. Such vaterites can be trapped in a transverse spin beam as described above, and are a good prospect for the transfer of transverse spin angular momentum. However, whether that desired transfer of angular momentum can be achieved depends on the orientation of the vaterites in the optical trap. It will be very useful to investigate this using computational modelling before proceeding to experiments.

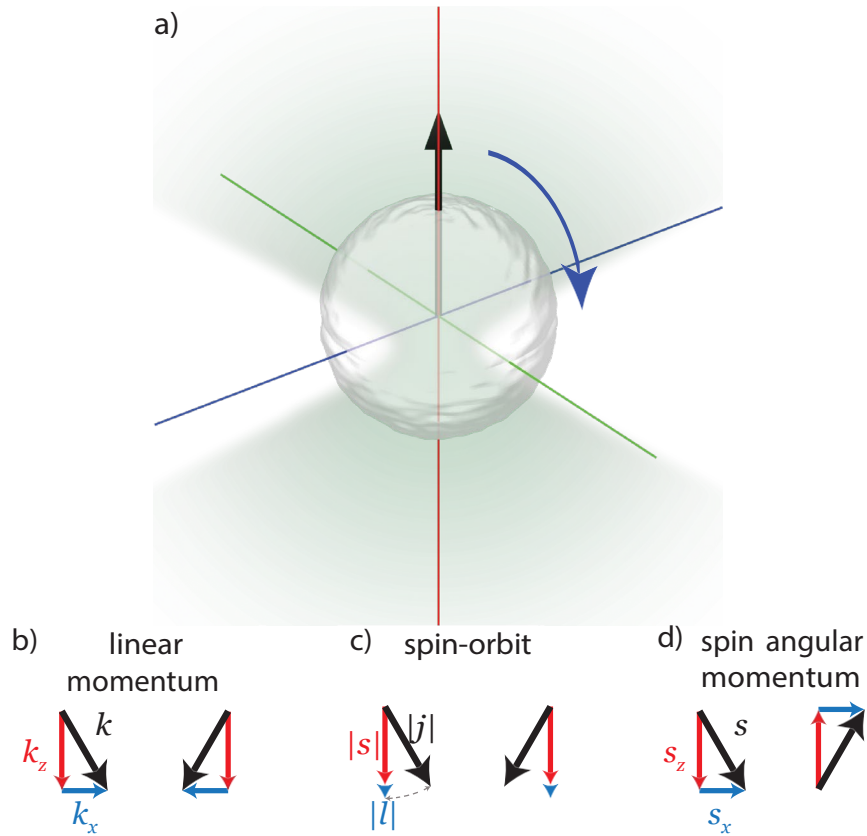

FIG. 1. a) Pictorial representation of a vaterite in a particle centered frame with optical axis represented by black arrow aligned with the beam and the transverse spin denoted by the magenta arrow. Momentum from two crossed focused beams: b) Since only the axial component of the local linear momentum density contributes to the total momentum flux of the beam, a converging beam has a lower momentum flux than a collimated beam of the same power. c) In the same way that the linear momentum flux is lower in a converging beam, the spin flux must also be lower in a converging beam. However, the total angular momentum flux remains the same, so an equal amount of orbital angular momentum must compensate for the reduced spin. d) If instead of combined two converging rays with equal spin, we combine two rays with opposite spin, the combined beam has zero axial spin flux, and has—unusually—non-zero transverse spin.

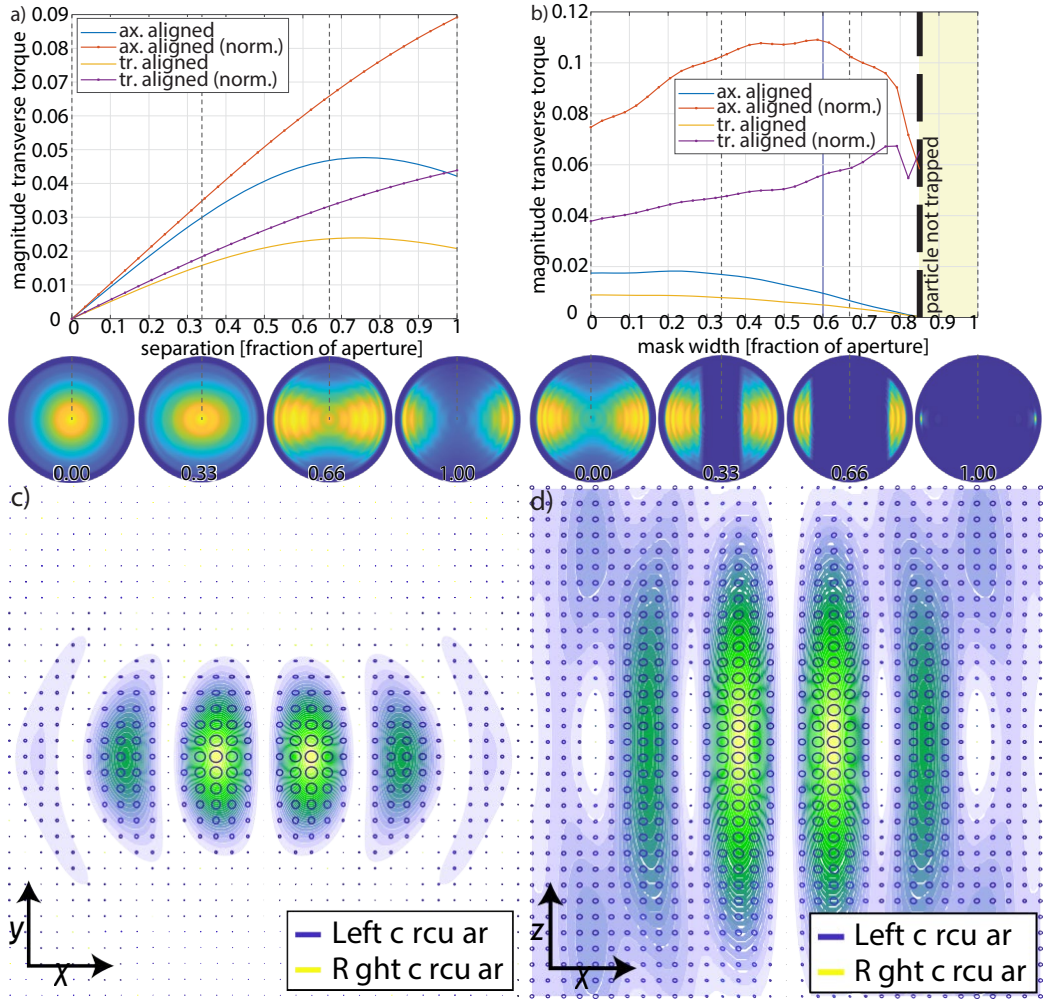

FIG. 2. a) Transverse Angular Momentum (TAM) as a function of beam separation. b) TAM as a function of mask thickness for a fixed beam separation. Transverse spin angular momentum density orthogonal in the c) *x, y* and d) *z, x* planes.

**COMPUTER SIMULATION OF CONTROLLED MOMENTUM TRANSFER**

We modelled the torque transfer to vaterite particles using the optical tweezers toolbox [35, 36]. We calculated the torque transferred to a 3.5 μm diameter vaterite at the stable axial trapping location with its optical axis aligned with the beam and orthogonal to the axis of transverse angular momentum (Supplementary Information, section "Modelling"). In Fig. 2a) and b), both the absolute and per photon torque increase for both alignments are depicted as the beams are initially separated. When the optical axis of the vaterite is orthogonal to the direction of the beam propagation the torque is about half that when aligned with the beam direction. At large separations, the torque per photon continues to increase whilst the reduction in radiance begins to dominate the contribution to the absolute torque. A mask of varying thickness was introduced between the two beams to cut overlapped regions of opposing spin angular momentum. We checked whether the torque per photon and the difference between the two alignments could be improved with this mask. As shown in Fig. 2b), a slight improvement in both the torque per photon and difference in alignments can be seen when the mask blocks more than 75% of the beams. The transverse spin angular momentum density of the composte beam is shown in Fig. 2c) focal plane and d) through the beam. Analysis of transverse spin angular momentum density in plane-wave interference has been shown previously [37].

There is a practical limit of improvement as there is a trade-off between the enhancement of per photon transfer and observed rotation as the transmitted light reduces. The mask fill level used in our experiment is estimated to be about 60% of the back aperture of the objective lens. Torque transfer per photon in our model is about $0.01\hbar$ per photon for both alignments, which is similar in magnitude to $0.02\hbar$ per photon found for axial torque transferred when the vaterite was trapped in a single circularly polarized beam [34]. Based on the results in Fig. 2 the angular velocity of the vaterite in a viscous fluid will be greatest when the optical axis is aligned with the beam and at most half that when it is orthogonal to it. For an aid to visualisation the transverse spin angular momentum density of the beam that results from the structuring is shown in Fig. 2c. As can be seen, the transverse spin component is present in a large portion of the beam near the focus.

**EXPERIMENTAL RESULTS**

To observe the transfer of transverse angular momentum we used a modified optical trapping apparatus (Supplementary information, Fig. S3). The key addition to the system are beam displacers to offset a single diagonally polarized beam into horizontal and vertical components. These components are then changed to orthogonal circular polarizations with a $\lambda/2$ waveplate. We determined the transfer of transverse spin angular momentum using spatially resolved Stokes measurements. The time-averaged spin angular momentum flux can be measured via the Stokes parameters over a spherical surface [27] is

$$\langle s \rangle = \frac{\int |E_+|^2 - |E_-|^2 \; \hat{r} \mathrm{d}\Omega}{\int |E_+|^2 + |E_-|^2 \; \mathrm{d}\Omega} \tag{3}$$

where $E_\pm$ is the right and left circularly polarized component of the light field, and $\hat{r}$ represents an element of the directional unit vector. The sampling of the back focal plane is equivalent to sampling the spherical surface where the objective lens approximately obeys the Abbe sine condition.

The orbital component of the angular momentum is more challenging as it requires determination of the phase gradient of light (Supplementary information, section "Difference between the Stokes measurement and optical torque").

Fig. 3a and 3b show the results from our simulations and measurements of spin angular momentum. Firstly, we see that when comparing the simulated and theoretical spin angular momentum (solid lines) we see that the simulation and theory produce a similar level of spin angular momentum transfer. The average spin angular momentum transfer in the simulation was about $0.12\,\hbar$ per photon. The spin angular momentum transfer was about $0.09\,\hbar$ per photon in the experiment. The oscillatory behaviour of the spin angular momentum predicted by the model was observed in the experiment as well. This arises due to the difference in the interaction with the beam as the particle rotates between the beam axis and the image plane. In a regular spin transfer experiment—where the particle rotates around a symmetric beam axis—a near constant transfer of spin angular momentum would be exhibited.

Interestingly, the total electromagnetic torque in the simulation does not match the spin angular momentum transfer. This is a clear signature of the presence of spin–orbit coupling. It also means that the spin angular momentum transfer measurement may not be used as a proxy for the total transfer of transverse angular momentum. Quantitative characterisation of the optical torque applied to the environment if the vaterite were to be used to drive a flow would require a full measurement of the transverse angular momentum. There is no equilibrium position under continuous rotation and thus a quantitative transverse spin measurement requires the unscattered beam spin angular momentum to be known as a reference.

In contrast to transverse rotation of nanoparticles, the transverse spinning motion of micron-size scale vaterite can be directly observed using several techniques. In addition to the Stokes measurement to determine the spin component of angular momentum, we can measure the waveplate nature of the particle as it spins using cross-polarizers (see supplementary material) and induce fluid flows to drive particles through the microscope imaging plane. Fig. 4 depicts two series of frames showing a naturally occurring small particle (likely debris from disintegrating vaterite) move through the focal plane on the left of the image, and then after a small gap in time (to omit frames with indistinct images), go back through the focal plane and behind the vaterite. By changing the parameters of the beam, e.g. balance in radiance between the two polarizations, position of the in-coupled light on the back focal plane, the rotation seems to be able to occur around multiple rotation axes. This opens the possibility of creating 3D shear flows with passive rotation of multiple micro-rotors. For examples, see Supplementary Movies.

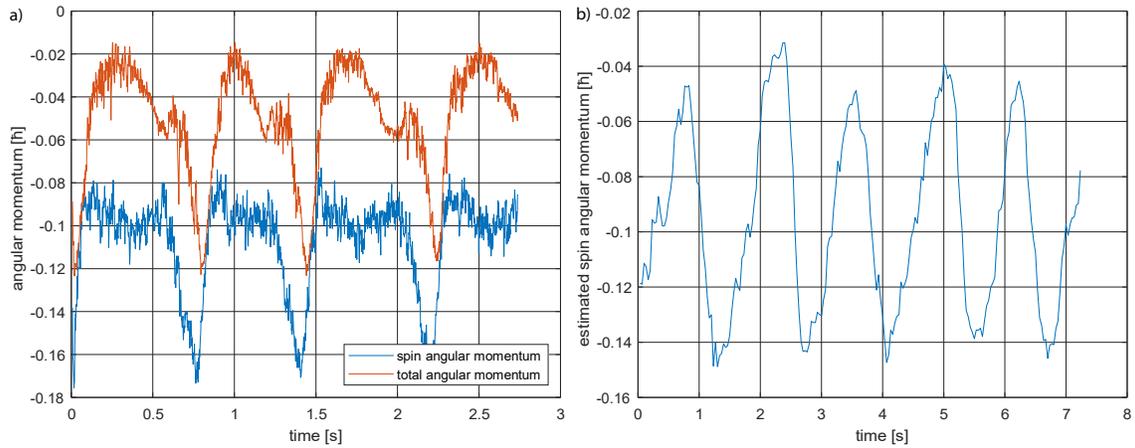

FIG. 3. Comparison of spin angular momentum determined from the simulation and experiment. a) Spin angular momentum and total angular momentum transferred to the simulated particle. b) Estimated spin angular momentum from the particle trapped in the experiment.

**DISCUSSION AND CONCLUSION**

We have demonstrated a novel controlled transfer of the transverse component of spin and orbital angular momentum to a microscopic vaterite particle. The spinning particle could generate a hydrodynamic flow field strong enough to drive another particle around it. Our modelling of the vaterite is consistent with our observations and provides good support for the changes in rotation rate as a function of orientation. We noted a small orbit of the particle around the beam as its orientation changed. These observations are consistent with the presence of spin–orbit coupling. Due to the generation of transverse angular momentum through interference, the details of the transfer of angular momentum to particles in the few-wavelengths size regime is much more complex than for nanoparticles and needs further investigation.

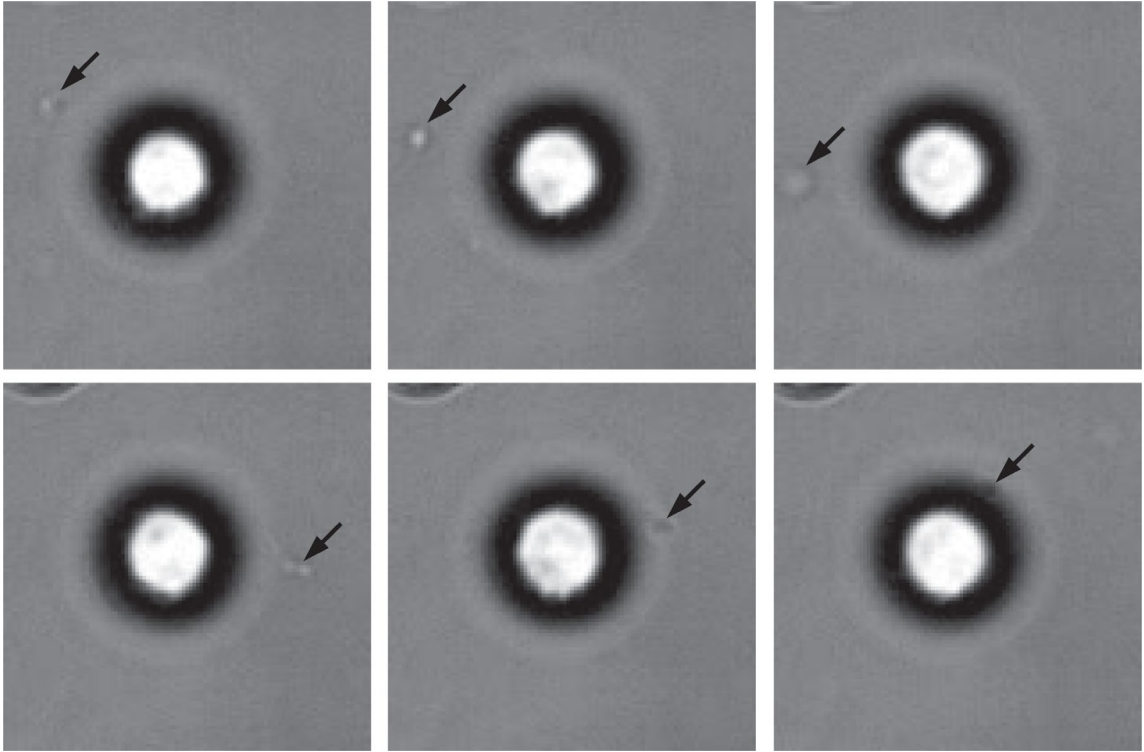

FIG. 4. Two sequences of frames from supplementary movie 1 at 10 frame intervals. In the top row, a small particle can be seen moving about the vaterite, changing its position relative to the image plane, first becoming brighter and then less distinct. The bottom row shows the opposite effect occurring on the other side of the vaterite, where the particle begins to darken and move behind the vaterite.

We believe that there are several potential future studies and experiments that can be performed using transfer of transverse angular momentum of light. This system provides new challenges in both transfer and measurement of electromagnetic torque transfer—the development of direct and general measurements of the total angular momentum would greatly improve our understanding of the interactions observed here and contribute the characterisation of other phenomena of optical torque transfer.

Transverse angular momentum transfer to probe particles would be highly advantageous in microscopic systems such as living cells as they are often very flat, and if we want to introduce a shear stress, it can be used to study mechanotransduction within the cell. Observation of light–matter interactions within optical traps gives us the ability to probe the mechanical effects of optical torque transfer around all principle axes.

Combined with existing methods for controlling, applying, and transferring spin and orbital angular momentum about the beam axis, transverse angular momentum offers the opportunity for the controlled application of full three-dimensional torques, with three independent orthogonal axes. Since the geometry of combining left and right circularly polarized beams at an angle to give transverse angular momentum also applies to beams carrying opposed orbital angular momentum, such as counter-helical optical vortices, a desired combination of both transverse spin and transverse orbital angular momentum can be produced.

This opens radically new opportunities in optical micromanipulation. For example, the three-axis rotation of birefringent microparticles or engineered particles driven by orbital angular momentum can be used in novel optically-driven microfluidic devices. Such three-dimensional control might also allow cooling of the rotational motion of an optically-trapped particle to the quantum ground state.

**ACKNOWLEDGEMENTS**

This research was supported by the Australian Research Council Discovery Project DP180101002. This research was also supported by the Australian Research Council

---

* a.stilgoe@uq.edu.au

[18] Andrea Aiello, Peter Banzer, Martin Neugebauer, and Gerd Leuchs. From transverse angular momentum to photonic wheels. *Nature Photonics*, 9(12):789–795, nov 2015.

[19] Martin Neugebauer, Thomas Bauer, Andrea Aiello, and Peter Banzer. Measuring the transverse spin density of light. *Physical Review Letters*, 114(6), feb 2015.

[20] T. A. Nieminen, A. B. Stilgoe, N. R. Heckenberg, and H. Rubinsztein-Dunlop. Angular momentum of a strongly focused Gaussian beam. *Journal of Optics A: Pure and Applied Optics*, 10:115005, 2008.

[21] Robert C. Devlin, Antonio Ambrosio, Noah A. Rubin, J. P. Balthasar Mueller, and Federico Capasso. Arbitrary spin-to–orbital angular momentum conversion of light. *Science*, 358(6365):896–901, nov 2017.

[22] Zengkai Shao, Jiangbo Zhu, Yujie Chen, Yanfeng Zhang, and Siyuan Yu. Spin-orbit interaction of light induced by transverse spin angular momentum engineering. *Nature Communications*, 9(1), mar 2018.

[23] Jörg S. Eismann, Peter Banzer, and Martin Neugebauer. Spin-orbit coupling affecting the evolution of transverse spin. *Physical Review Research*, 1(3), dec 2019.

[24] Ankan Bag, Martin Neugebauer, Paweł Woźniak, Gerd Leuchs, and Peter Banzer. Transverse kerker scattering for angstrom localization of nanoparticles. *Physical Review Letters*, 121(19):193902, nov 2018.

[25] J. Humblet. Sur le moment d'impulsion d'une onde électromagnétique. *Physica*, 10(7):585–603, jul 1943.

[26] J. M. Jauch and F. Rohrlich. *The Theory of Photons and Electrons*. Springer, Berlin, 1976.

[27] J. H. Crichton and P. L. Marston. The measurable distinction between the spin and orbital angular momenta of electromagnetic radiation. *Electronic Journal of Differential Equations*, Conf. 04:37–50, 2000.
18